\documentclass[showpacs, twocolumn]{revtex4-1}

\usepackage{graphicx}
\usepackage{amsmath}

\begin{document}
\title{Thermodynamics of a two-dimensional dipolar Bose gas with correlated disorder in the roton regime}

\author{Abdel\^{a}ali Boudjem\^{a}a}

\affiliation{Department of Physics, Faculty of Sciences, Hassiba Benbouali University of Chlef P.O. Box 151, 02000, Ouled Fares, Chlef, Algeria}

\email {a.boudjemaa@univ-chlef.dz}

\begin{abstract}
We sudy the impact of a weak random potential with a Gaussian correlation function on the thermodynamics of a two-dimensional (2D) dipolar bosonic gas. 
Analytical expressions for the quantum depletion, anomalous density, the ground state energy, the equation of state and the sound velocity  are 
derived in the roton regime within the framework of the Bogoliubov theory. 
Surprisingly, we find that the condensate depletion and the anomalous density are comparable.
The structure factor and the superfluid fraction are also obtained analytically and numerically.
We show that these quantities acquire dramatically modified profiles when the roton is close to zero yielding the transition to an unusual quantum state.
\end{abstract}


\maketitle

\section{Introduction} \label{lntrod}

In past years, dipolar Bose gases have attracted a great deal of interest both theoretically and experimentally
\cite {Baranov, Pfau, Carr, Pupillo2012}. 
The dipole-dipole interaction (DDI), which is long-range and anisotropic, plays a crucial role when compared to the short-range isotropic interaction.
These special features of the DDI lead to the observation of novel phenomena in ultracold dipolar gases. 
Interesting structural properties emerge in such systems is the presence of the low-lying roton minimum in
the excitation spectrum \cite {gora, boudjGora} and the possibility of the crystallization of solid bubble
into a lattice superstructure, resulting in a global supersolid phase \cite {prok, boudjGora, petGora}.
However, such supersolids require a dense regime with several particles within the interaction range, which can be
difficult to achieve. 
The appearance of the roton-maxon character in the excitation spectrum of pancake dipolar condensates has been predicted first
by  Santos et\textit {al}.\cite{gora}  where it has been shown that upon further decreasing the confining trap frequency 
the roton energy drops to zero triggering a dynamical instability.  
Since that time, there has been a number of recent theoretical studies proposing schemes to detect rotons and characterize their
effects on the properties of dipolar Bose-Einstein condensate (BEC) (e.g., see \cite{Uwe, Ron, bohn1, Lu, Wils, Mart, Tic, Nath, Biss}). Experimentally, rotons will likely be realized in pancake
shaped traps, and the trap itself plays a very fundamental role in the nature of the rotons that emerge \cite {Ron}.

The roton-maxon character of the Bogoliubov spectrum was originally observed in ${}^4$He superfluid 
and arises due to the strong isotropic repulsion between the atoms in the liquid  \cite{Land, Land1,Feym, Feym1}.
However, the nature of the roton in the context of quasi-2D dipolar BECs is radically different. In these dilute systems, the roton is originated from the anisotropy of the DDI interaction. 
Other recent studies showed also the formation of the roton-maxon excitation spectrum for a weakly correlated Bose gas of dipolar excitons in a semiconductor layer \cite{Fedo}.
It was found that the presence of the roton minimum in 2D dipolar bosons leads to reduce the condensed fraction even at zero temperature \cite{mora,bush,Astr}.
Finite temperature Monte Carlo simulations \cite{prok3} have revealed that the rotonization of the spectrum can decrease the Kosterlitz-Thouless (KT) superfluid
transition temperature. 
As for the pancaked dipolar BEC, it has been pointed out that the roton modes serve to change the sign of the anomalous density near the trap center for large values of DDI \cite {Blak2}.

On the other hand, disorder has been observed to cause a dramatic influence on a BEC and has sparked immense interest recently \cite {San, Deis, Ale}.
Particles moving in a disordered environment may open fascinating prospects for the observation of non-trivial quantum phases.
Among the striking features resulting from the presence of a disordered external potential is the existence of an insulating phase called “Bose glass” \cite {Giam, Fisher}
and the occurrence of Anderson localization \cite {And} in non-interacting systems. Special attention has been paid to this phenomenon \cite {Sanch, Asp, Asp1} in recent years.

One of the first attempts to study the so-called dirty boson problem was introduced by Huang and Meng in 1992 \cite{Huang}. 
Most recently, this approach which  is modeled by a uniform random distribution of quenched impurities and
based on the Bogoliubov theory, has been extended to the case of a 3D dipolar BEC with both uncorrelated 
\cite{AxelM} and correlated \cite{Boudj,Boudj1} disorder potential.
The main result emerging from these studies is that the anisotropy of the two-particle DDI is passed on to the superfluidity 
i.e. the superfluid density acquires a characteristic direction dependence \cite{AxelM, Boudj,Boudj1,Axel1,Axel2}. 
This peculiar phenomenon, which is not present at zero temperature in the absence of disorder, contributes to a deeper understanding of the localization phenomenon. 
In the case of correlated disorder, it was found that both condensate and superfluid depletions decrease with increasing the disorder correlation length. 
This picture holds also in a quasi-2D dipolar BEC with $\delta$-correlated disorder \cite{Boudj2}. 

In this paper, we investigate the thermodynamics of a quasi-2D dipolar Bose gas subjected to a weak random potential with Gaussian correlation by using the Huang-Meng-Bogoliubov theory.
The most important feature of the Gaussian-correlated disorder is that it renders the macroscopic wave function of the BEC insensitive to the disorder strength, but instead depends 
on the disorder averaged over the correlation length \cite {Mich, Sanch1}.
The presence of the Gaussian-correlated disorder potential in 1D Bose gas with 
short range potential can undergo a finite-temperature phase transition between two distinct states: fluid and insulator \cite{Ale}.
Furthermore, random potential with Gaussian correlation function constitutes an efficient tool to describe the Bose glass phase in 2D and 3D BEC with conatct interaction 
in both lattice and continuum models \cite {Mich, Sal, Allard}. 
In addition, in 2D Bose gas with DDI, one can expect that such a Gaussian-correlated random potential may yield the transition into the superglass state \cite{Boudj1,Boudj2}.

The rest of the paper is organized as follows. In Sec.\ref {model}, we review the main features of the Huang-Meng-Bogoliubov theory
of dipolar dilute Bose gas in a general disorder potential.
In Sec.\ref{FlTh},  we apply this approach to derive analytical expressions for the condensate fluctuations and
thermodynamic quantities such as the chemical potential, the ground state energy and the sound velocity for Gaussian correlated disorder potential in the roton regime.
We analyze the behavior of noncondensed and anomalous densities in terms of the temperature and the interaction strength.
Likewise, we calculate the corrections to the sound velocity due to the correlated disorder. 
In Sec.\ref{SFSLD}, we deeply investigate the properties of the structure factor and the superfluid fraction with respect of the system parameters.
Finally, we discuss and summarize our results in Sec.\ref{Concl}.

\section{The Model} \label{model}

We consider a dilute Bose-condensed gas of dipolar bosons in an external random potential $U({\bf r})$. 
These particles can be confined to quasi-2D, by means of an external harmonic potential
in the direction perpendicular to the motion (pancake geometry) and all dipoles are aligned perpendicularly to the plane of their translational
motion, by means of a strong electric (or magnetic) field. 
In this quasi-2D geometry, at large interparticle separations $r$ the interaction potential is  
$V(r) = d^2/ r^3=\hbar^2r_*/mr^3$, with $d$ being the dipole moment, $m$ the particle mass, and $r_*=md^2/\hbar^2$ is the characteristic dipole-dipole distance.
From now on, we assume that $U({\bf r})$ is a weak external potential with vanishing ensemble averages $\langle U({\bf r})\rangle=0$
and a finite correlation of the form $\langle U({\bf r}) U({\bf r'})\rangle=R ({\bf r},{\bf r'})$.


In the ultracold limit where the particle momenta satisfy the inequality $kr_*\ll1$, the scattering amplitude is given by (see e.g. \cite{boudjGora})
\begin{equation}\label{ampl} 
 f({\bf k},{\bf k'})=g(1-C\vert {\bf k}-{\bf k'}\vert),
\end{equation}
where the 2D short-range coupling constant is $g=g_{3D}/\sqrt{2}l_0$ and $C =2\pi \hbar^2r_*/mg=2\pi d^2/g$.
Employing this result in the secondly quantized Hamiltonian, we obtain

\begin{align}\label{he3}
&\hat H\!\!=\!\!\sum_{\bf k}\!E_k\hat a^\dagger_{\bf k}\hat a_{\bf k}\! +\!\frac{1}{S}\!\!\sum_{\bf k,\vec p} \! U_{\bf k\!-\!\bf p} \hat a^\dagger_{\bf k} \hat a_{\bf p} \\ \nonumber
&+\!\frac{g}{2S}\!\!\sum_{\bf k,\bf q,\bf p}\!\!
(1\!\!-\!C\vert {\bf q\!-\!\bf p}\vert)\hat a^\dagger_{\bf k\!+\!\bf q} \hat a^\dagger_{\bf k\!-\!\bf q}\hat a_{\bf k\!+\!\bf p}\hat a_{\bf k\!-\!\bf p} ,
\end{align}
where $S$ is the surface area, $E_k=\hbar^2k^2/2m$, and $\hat a_{\bf k}^\dagger$, $\hat a_{\bf k}$ are the creation and annihilation operators of particles.
At zero temperature there is a true BEC in 2D, and we may use the standard Bogoliubov approach.
Assuming the weakly interacting regime where $mg/2\pi\hbar^2\ll 1$ and $r_*\ll \xi$, with $\xi=\hbar/\sqrt{mng}$ being the healing length, 
we may reduce the Hamiltonian (\ref{he3}) to a bilinear form, using the homogeneous Bogoliubov transformation \cite{Huang}
\begin{equation}\label {trans}
 \hat a_{\bf k}= u_k \hat b_{\bf k}-v_k \hat b^\dagger_{-\bf k}-\beta_{\bf k},  \qquad \hat a^\dagger_{\bf k}= u_k \hat b^\dagger_{\bf k}-v_k \hat b_{-\bf k}-\beta_{\bf k}^*,
\end{equation}
where $\hat b^\dagger_{\bf k}$ and $\hat b_{\bf k}$ are operators of elementary excitations.\\
The Bogoliubov functions $ u_k,v_k$ are expressed in a standard way: $ u_k,v_k=(\sqrt{\varepsilon_k/E_k}\pm\sqrt{E_k/\varepsilon_k})/2$, 
$\beta_{\bf k}=\sqrt{n/S} U_k E_k/\varepsilon_k^2$, and  the Bogoliubov excitation energy is given by 
$\varepsilon_k=\sqrt{E_k^{2}+2\mu E_k(1-Ck)}$ with $\mu=ng$ being the zeroth order chemical potential.
If $C\leq (\sqrt{8}/3)\xi$, $\varepsilon_k$ is a monotonic function of $k$.
However, it shows a roton-maxon structure for the constant $C$ in the interval $(\sqrt{8}/3)\xi\leq C\leq\xi$.
It is then convenient to represent $\varepsilon_k$ in the form \cite{boudjGora}:
\begin{equation}\label{spec} 
\varepsilon_k= \frac{\hbar^2 k}{2m}\sqrt{ (k-k_r)^2 +k_{\Delta}^2},
\end{equation}
where $k_r=2C/\xi^2$ and $k_{\Delta}=\sqrt{4/\xi^2-k_r^2}$.
If the roton is close to zero, then $k_r$ is the position of the roton, and 
$\Delta\!=\!\hbar^2 k_rk_{\Delta}/2m$, 
is the height of the roton minimum.
For $C=\xi$ the roton minimum touches zero, and at larger $C$ the uniform Bose condensate becomes dynamically unstable.
We see from Eq.(\ref {spec}) that the spectrum energy is independent of the random potential which means that the standard Bogoliubov theory (zeroth order in perturbation theory)
cannot predict any change in the excitations dispersion. 

The diagonal form of the Hamiltonian of the dirty dipolar Bose gas (\ref{he3}) can be written as
\begin{equation}\label{DHami} 
\hat H = E+\sum\limits_{\bf k} \varepsilon_k\hat b^\dagger_{\bf k}\hat b_{\bf k},
\end{equation}
where $E=E_0+ E'+ E_R$ with $E_0= S g n^2/2 $ and 
\begin{equation}\label{Fenergy} 
E'=\frac{1}{2}\sum\limits_{\bf k} [\varepsilon_k -E_k-ng(1-Ck)],
\end{equation}
 being the ground-state energy correction due to quantum fluctuations. 
\begin{equation}\label{Renergy} 
E_R=-\sum\limits_{\bf k} n\langle |U_k|^2\rangle \frac{ E_k}{\varepsilon_k^2} =-\sum\limits_{\bf k} n R_k \frac{ E_k}{\varepsilon_k^2},
\end{equation}
gives the correction to the ground-state energy due to the external random potential.

The noncondensed and the anomalous densities of a disordered BEC are defined, respectively as \cite{Boudj, Boudj1}:
\begin{equation}\label {dep}
\tilde{n}=\sum_{\bf k} \langle\hat a^\dagger_{\bf k}\hat a_{\bf k}\rangle=\tilde{n}' +n_R,
\end{equation}
and 
\begin{equation}\label {anom}
\tilde{m}=\sum_{\bf k} \langle\hat a_{\bf k}\hat a_{-\bf k}\rangle=\tilde{m}'+n_R,
\end{equation}
where 
\begin{equation}\label {dep1}
\tilde{n}'=\frac{1}{2}\int \frac{d^2k} {(2\pi)^2} \left[\frac{E_k+ng(1-Ck)} {\varepsilon_k}\coth\left(\frac{\varepsilon_k}{2T}\right)-1\right],
\end{equation}
\begin{equation}\label {anom1}
\tilde{m}'=-\frac{1}{2}\int \frac{d^2k} {(2\pi)^2} \frac{ng(1-Ck)} {\varepsilon_k}\coth\left(\frac{\varepsilon_k}{2T}\right),
\end{equation}
are the condensate depletion and anomalous density due to quantum fluctuations, respectively.
\begin{equation}\label {depdis}
n_R=\frac{1}{S}\sum\limits_{\bf k} \langle |\beta_k|^2\rangle=\frac{1}{2}\int \frac{d^2k} {(2\pi)^2}  nR_k \frac{ E_k^2}{\varepsilon_k^4},
\end{equation}
is the density correction due  to the external random potential.

\section{Fluctuations and Thermodynamics} \label{FlTh}
In what follows, we consider the case of a weak external random potential with Gaussian correlation which can be written in the momentum space as
$R({\bf k}) = R_0 \, e^{-\sigma^2k^2/2}$, where $R_0$ with dimension (energy) $^2$ $\times$ (length)$^2$ and $\sigma$ characterize
the strength and the correlation length of the disorder, respectively.
Indeed, this type of disorder potential makes our study substantially more detailed, general and rigorous
since uncorrelated random potentials are usually crude approximations of realistic disorder, for which $\sigma$ can be significantly large.

Assuming now that the roton is close to zero and the roton energy is $\Delta\ll \mu$, we have the coefficient $C$  close to $\xi$, and $k_r\simeq 2/\xi$. 
Then, using Eq.(\ref{depdis}) for the contribution of momenta near the roton minimum at $T=0$, we obtain:
\begin{equation} \label{depdis1}
\frac{n_R}{n}=\frac{m g}{4 \hbar^2} \left(\frac{2\mu}{\Delta}\right)^3 R\, e^{-2\sigma^2/\xi^2};\,\,\,\,\,\Delta\ll \mu,
\end{equation}
where $R=R_0/ng^2$ is a dimensionless disorder strength.\\
For $\sigma/\xi\rightarrow 0$, the disorder fluctuation (\ref{depdis1}) reduces to that of dipolar BEC with $\delta$-correlated disorder \cite{Boudj2}.\\

Integrals  (\ref{dep1}) and  (\ref{anom1}) are logarithmically divergent at large momenta because of the dipolar contribution to the interaction strength $-gCk$ \cite{boudjGora}.
To overcome this problem, we can resort to a high momentum cut-off $1/r_*$. 
Inserting the resulting expressions in (\ref{dep}) and  (\ref{anom}), we obtain for the condensate depletion and the anomalous fraction:
\begin{equation}    \label{Fluc}
\frac{\tilde{n}}{n} \approx \frac{\tilde{m}}{n} \simeq \frac{mg}{\pi\hbar^2}\left[\ln\left(\frac{2\mu}{\Delta}\zeta\right)+\frac{\pi}{4} \left(\frac{2\mu}{\Delta}\right)^3 R\,e^{-2\sigma^2/\xi^2}\right],
\end{equation}
where $\zeta=\sqrt{2\pi\hbar^2/e^2 mg}$.\\
The leading term in Eq.(\ref{Fluc}) was first obtained in our recent work \cite {boudjGora}, while the second term represents the disorder correction to the noncondensate and anomalous fractions.
For $\sigma/\xi\gg1$, the disorder effects become negligible and hence, the condensed fraction takes the form $n_c/n\simeq1- (mg/\pi\hbar^2)\ln (2\mu\zeta/\Delta)$.
Furthermore, equation (\ref{Fluc}) clearly shows that the anomalous density and the condensate depletion are comparable in the roton branch. 

\begin{figure}[htb1] 
\includegraphics[scale=0.8]{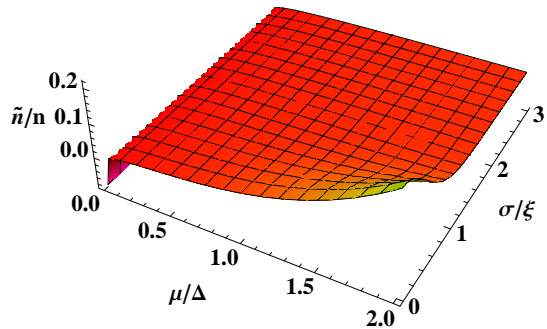}
\caption {Quantum depletion of  dirty dipolar condensate, as a function of $\mu/\Delta$  and $\sigma/\xi$ for $mg/4\pi\hbar^2=0.01$
and $R=0.1$. }
\label{depl}
\end{figure}

Figure.\ref{depl} shows that in the absence of the random external potential i.e. $R=0$, the noncondensed fraction grows logarithmically (see Eq. (\ref{dep1}))  
when the roton energy $\Delta$ goes to zero yielding the transition to a supersolid state \cite {prok, boudjGora, petGora}.
In the presence of the disorder potential the ratio of the correlation length and the healing length $\sigma/\xi$ decreases the condensate depletion according to the function $e^{-2\sigma^2/\xi^2}$.

The Bogoliubov approach assumes that the condensate depletion should be small. We thus conclude from Eq.~(\ref{dep1}) that at $T=0$ and for the roton minimum close to zero, 
the validity of the Bogoliubov approach is guaranteed by the inequalities 
$\left(mg/\pi\hbar^2\right)\ln (2\mu/\Delta\zeta) \ll 1$, and $\left(mg/\hbar^2\right) (2\mu/\Delta)^3 R\,e^{-2\sigma^2/\xi^2} \ll 1$.

However, the situation changes in the calculation of the correction to the ground-state energy due to the external random potential.
When the roton minimum is approaching to zero, we get from (\ref{Renergy})
\begin{equation}   \label{Renergy1}
\frac{E_R}{E_0}=-\frac{2mg}{\hbar^2} \left(\frac{2\mu}{\Delta}\right) R\, e^{-2\sigma^2/\xi^2};\,\,\,\,\,\,\Delta\ll \mu.
\end{equation}
Equation (\ref {Renergy1}) shows that $E_R$ linearly depends on $ng/\Delta$, and decreases with increasing  $\sigma/\xi$. 
Furthermore, the correction (\ref {Renergy1}) is negative which means that the random potential leads to lower the total energy of the system.\\
The correction to the chemical potential due to disorder effects is then obtained easily through $\partial E_R/\partial N$
\begin{equation}    \label{deltamu}
\frac{\mu_R}{\mu}= -\frac{mg}{\hbar^2} \left(\frac{2\mu}{\Delta}\right)^3 R\,e^{-2\sigma^2/\xi^2};\,\,\,\Delta\ll \mu.
\end{equation} 

The shift of the ground-state energy due to quantum fluctuations can be given as 
\begin{equation}   \label{energy1}
\frac{E'}{E_0}\simeq 1+\frac{2mg}{\pi\hbar^2}+\frac{2mg}{\pi\hbar^2}\ln\left(\frac{2\mu}{\Delta}\right);\,\,\,\,\,\,\Delta\ll \mu.
\end{equation}
Note that quantum fluctuations correction to the chemical potential can be calculated straightforwardly using $\partial E'/\partial N$ (see e.g. \cite {boudjGora}).

The correction to the sound velocity can be simply calculated via $mc_s^2 = n \partial \mu/\partial n $ \cite {boudjGora,boudj2015, Lev} as
\begin{align}    \label{sound}
\frac{c_s^2}{c_{s0}^2} &= 1+\frac{mg}{\pi\hbar^2}\left[2\ln\left(\frac{2\mu}{\Delta}\right)+\left(\frac{2\mu}{\Delta}\right)^2\right] \\\nonumber
&+\frac{mg}{\hbar^2} \left[ \left(\frac{2\sigma^2}{\xi^2}+\frac{3}{2}\right) \left(\frac{2\mu}{\Delta}\right)^3-\frac{3}{2}\left(\frac{2\mu}{\Delta}\right)^5\right] R\,e^{-2\sigma^2/\xi^2},
\end{align} 
where $c_{s0}=\sqrt{\mu/m}$ is the zeroth order sound velocity. 
The second and the third terms originate from quantum fluctuations while the last term comes from the disorder contribution. 
For $\sigma/\xi\rightarrow 0$, the sound velocity (\ref{sound}) becomes identical to that obtained in quasi-2D dipolar BEC with delta-correlated disorder \cite {Boudj2}.
\begin{figure}
\centerline{
\includegraphics[scale=0.8]{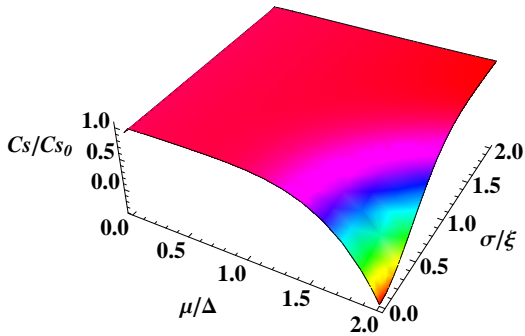}}
\caption {(Color online) Sound velocity as a function of $\mu/\Delta$ and $\sigma/\xi$. Parameters are the same as in Fig.\ref{depl}.}
\label{speed}
\end{figure} 
Equation (\ref{sound}) shows that the main correction to the sound velocity due to the disorder potential is negative $\sim-(2\mu/\Delta)^5\,R\,e^{-2\sigma^2/\xi^2}$.
We see from Fig.(\ref{speed}) that for $\sigma \ll \xi$, $c_s$ is practically constant in the range $0<\Delta\leq \mu$, while it 
reduces and vanishes at $\Delta\simeq \mu/2$. 
This value can be changed with increasing or decreasing the disorder strength $R$. 
For $\sigma >\xi$, $c_s$ rises with rising $\mu/\Delta$. \\
It is worth stressing that, in 3D disordered BECs with a pure contact interaction, the sound velocity has been calculated with different approaches leading to different predictions. 
For instance, standard perturbation theory predicts an increased $c_s$ in Bose gas with an uncorrelated disorder \cite{Gior, Fal}.
On the other hand, the extended Bogoliubov approach developed in \cite{Gaul, Gaul1} and the mean field theory of Ref \cite{Yuk} provide a reduced sound velocity.

\section{Structure factor and Superfluid fraction}\label{SFSLD}

As is known in an infinite uniform 2D fluid, thermal fluctuations at any nonzero temperature are strong enough
to destroy the fully ordered state associated with BEC, but are not strong enough to suppress superfluidity in an interacting system at low, but non-zero
temperatures \cite{merm, hoh}. 
However, according to KT \cite{KT}, such a transition is associated with the unbinding of vortex pairs or
quasi-long-range order. Below the KT transition temperature, a 2D Bose gas (liquid) is characterized
by the presence of a “quasicondensate” \cite{pop, petr1, boudj2012}.
In this quasicondensate, the phase coherence governs only regimes of a size smaller than
the size of the condensate, characterized by the coherence length $l_\phi$\cite {petr1, boudj2012}.
Thermodynamic properties, excitations, structure factor and correlation properties on a distance scale smaller than $l_{\phi}$ are the same as in the case of a true BEC. 
Moreover, for realistic parameters of quantum gases, $l_{\phi}$ exceeds the size of the system \cite{GPS}, so that one can employ the ordinary BEC theory. 
Therefore, the correction to the condensate depletion and thermodynamic quantities due to thermal fluctuations turns out to be given as $(2mg/\hbar^2)T/\Delta$ \cite{boudjGora}.

\subsection{Structure factor}
The static structure factor which is the Fourier transform of the density-density correlation function is defined through the relation
\begin{equation}   \label{SF}
S({\bf k})= \frac{1}{N}\langle \hat n_k \hat n^\dagger_k\rangle,
\end{equation}
where
$$\hat n_k= \sqrt{N} \left(\hat a_k^\dagger +\hat a_{-k}\right) + \sum_{q\neq 0} \hat a_{k+q}^\dagger \hat a_q.$$
Applying the Bogoliubov transformation (\ref{trans}) we obtain for the static structure factor
\begin{align}   \label{SF1}
S({\bf k})=\frac{E_k}{\varepsilon_k} \text {coth} \left(\frac{\varepsilon_k}{2T}\right) +4\langle \beta_k^2\rangle.
\end{align}
The central value $S(0)$ is given as
\begin{equation} \label{SF2}
S(0)=\frac{T}{\mu}+\frac{R_0}{ng^2},
\end{equation}
which is independent from the dipolar force.
For large momenta, the static structure factor approaches unity.
The above expressions show that the external random potential leads to an increase of the structure factor. 
Moreover, we see that $S(k)$ remains finite even at $T=0$ due to the presence of the disorder potential which makes the compressibility of the gas finite. 
\begin{figure}
\centerline{
\includegraphics[scale=0.58]{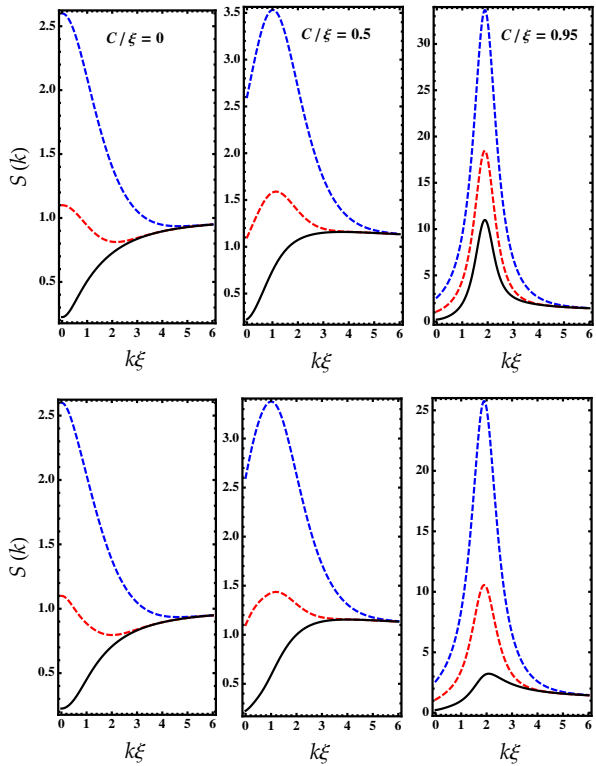}}
\caption {(Color online) Static structure factor from Eq.(\ref{SF1}) as a function of dimensionless variable $k\xi$ for different values of temperature and dipolar interaction.
Parameters are: $R=0.1$, $\sigma/\xi=0.2$ (top panels) and $\sigma/\xi=1.2$ (bottom panels). 
Blue dashed lines: $T=3\mu/2$, red dashed lines: $T=\mu/2$ and black solid lines: $T=\mu/16$.}
\label{SFF}
\end{figure} 

On the other hand, the static structure factor exhibits a strong dependence on the temperature and the DDI as is clearly seen from Fig.\ref{SFF}. 
At small value of the roton gap i.e. ($C\sim \xi$), $S(k)$ grows significantly and has a peak around $k= 1/\xi$ indicating that the thermal fluctuations of the density becomes important 
in the roton regime (see the last row of Fig. \ref{SFF}). Therefore, the system envisages a transition to a new quantum phase.
Note that this reduction and/or suppression of the BEC coherence has been confirmed experimentally in 2D Bose gas with correlated-disorder \cite{Allard}.
In the absence of the DDI, thermal effects are important at small $k$ i.e. in the phonon branch (see the first row of Fig.\ref{SFF}).\\
For a large disorder correlation i.e. $\sigma \geq \xi$, the structure factor decreases even when the roton minimum is close to zero leading to small density fluctuations (see bottom panels of Fig.\ref{SFF}).

\subsection{Superfluid fraction} 
In the context of the liquid helium, it has been shown that the position of the roton minimum influences the phenomenon of superfluidity \cite{Noz, bali}.
Here we look how the interplay of the rotonization and external disorder potenial can affect the superfluid fraction of a quasi-2D Bose gas with DDI.

The superfluid fraction $n_s/n$ can be found from the normal fraction $n_n/n$ which is determined by the transverse current-current
correlator $n_s/n =1-n_n/n$. We apply a Galilean boost with the total momentum of the moving system ${\bf \hat P_v}={\bf \hat P}+mv N$, where 
${\bf \hat P}=\sum_k \hbar {\bf k}\,\hat a^\dagger_{\bf k} \hat a_{\bf k}$ and $v$ is the liquid velocity. In the $d$-dimensional case, the superfluid fraction reads
\begin{equation}   \label{sup}
 \begin{split}
 \frac{n_s}{n}= 1-\frac{2}{dTn} \int \frac{d^dk}{(2\pi)^d} \left[\frac{E_k}{4 \text {sinh}^2 (\varepsilon_k/2T)} \right. \\+
\left. \frac{n R_k E_k^2}{\varepsilon_k^3}  \text {coth} \left(\frac{\varepsilon_k}{2T}\right) \right].
\end{split}
\end{equation}
At very low temperature we can put $\text {coth}(\varepsilon_k/2T)=2T/\varepsilon_k$. Thus, Eq. (\ref{sup}) reduces to   
\begin{equation}  \label{sup1}
 \frac{n_s}{n}= 1-\frac{4}{d}\frac{n_R}{n} -\frac{2}{dTn}\int \frac{d^dk}{(2\pi)^d} \left[ \frac{E_k}{4 \text {sinh}^2 (\varepsilon_k/2T)}\right]. 
\end{equation}
Equation.(\ref {sup1}) clearly shows that the ratio between the normal fluid density and the corresponding condensate depletion increases to 2 in 2D and to 4 in 1D, 
in contrast to the familiar 4/3 in 3D geometry obtained earlier in \cite {Huang, Gior, Lopa}. 
Remarkably, the superfluid density (\ref{sup1}) is a scalar quantity contrary to the 3D case where it has been
found that $n_s$ is a tensorial quantity\cite{AxelM, Boudj, Boudj1} due to the anisotropy of the DDI.

Assuming now that the roton minimum is close to zero, then the momenta near the roton minimum are the most important, this yields at $T=0$:
\begin{equation}   \label{super} 
\frac{n_s}{n}= 1-\frac{mg}{2\hbar^2} \left(\frac{2\mu}{\Delta}\right)^3 R\,e^{-2\sigma^2/\xi^2};\,\,\,\,\,\,\Delta\ll \mu. 
\end{equation}
This equation shows that for $\sigma/\xi\gg1$, $n_s\sim n$ in contrast to $n_c$ where this latter remains small even for $\sigma/\xi\gg1$
owing to the quantum fluctuation described by the logarithmic term. 

Now we turn to analyze numerically the normal fraction of the superfluid as a function of the ratio $\sigma/\xi$ 
and the strength of disorder for different positions of the roton minimum.

\begin{figure}
\centerline{
\includegraphics[scale=0.7]{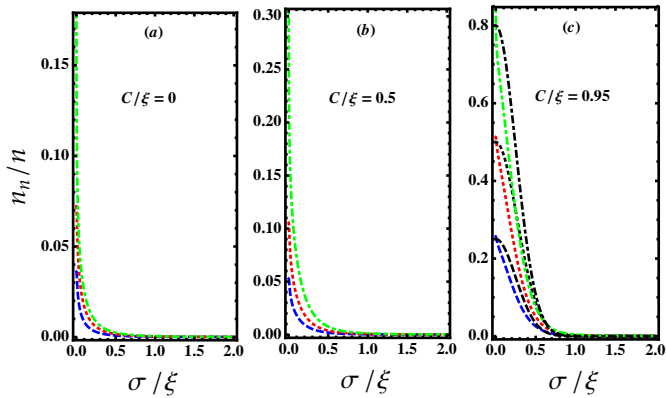}}
\caption {(Color online) Normal fraction from Eq.(\ref {sup}), as a function of $\sigma/\xi$ for $R=0.05$ (Blue dashed), $R=0.1$ (Red dotted) and $R=0.15$ (Green dotdashed). 
Parameters are: $T/\mu=0.2$ and $mg/4\pi\hbar^2=0.01$. Black lines represent analytical solutions.}
\label{22}
\end{figure} 
Figure.\ref{22} depictes that  for $\sigma \geq\xi$, the normal fraction vanishes  and thus, the system becomes completely superfluid
for any value of the disorder strength and the roton position. The reason is that  when the healing length of the BEC is
smaller than the correlation length of the disorder potential, the kinetic energy term is small and the BEC density simply follows the spatial
modulations of the potential and hence, the condensed particles will not localize \cite{Sanch1}. This result excellently coincides with our analytical predictions (\ref{super}).
Whereas, for $\sigma <\xi $, $n_n/n$ is increasing with $R$ and $C/\xi$. 
One can observe from the same figure that when the roton minimum is very close to zero ($C\sim \xi$) and for a large value of $R$, the normal fraction is significant
which makes it possible to destroy superfluidity even at very low temperature (see Fig.\ref{22}.c).
This is attributed to the fact that the particles are localized in the respective minima of the external random potential 
and thus form distributed randomly obstacles for the motion of the superfluid. 
However this localization is different from Anderson localization of Bogoliubov quasiparticles observed by Lugan et \textit {al}. \cite{Lug, Gaul3}. 
The Bogoliubov quasiparticles experience a randomness mediated by the inhomogeneous condensate background, which responds nonlinearly and nonlocally 
to an effective potential that is different from the usual bare disorder \cite {Gaul1,Lug,  Mul, Gaul3}.
Therefore, the localization properties are changed compared to bare particles although the general symmetry class is the same \cite {Mul, Gaul3}.

\section{Conclusions} \label{Concl}

We have investigated a dilute 2D dipolar Bose gas with dipoles oriented perpendicularly to the plane 
subjected to a weak Gaussian correlated disorder potential in the roton regime.
Using the  Bogoliubov approach, we have derived analytical expressions for the condensate depletion, the ground state energy, the equation of state, the sound velocity
the structure factor and the superfluid fraction.
Our analysis signifies that in the limit $\sigma/\xi\rightarrow 0$, the disorder potential strongly enhances the fluctuations and the thermodynamic quantities.
This may lead to the transition of a non-trivial quantum phase (disordered supersolid state). 
The qualitative study of the phase diagram of such a state requires either a non-perturbative approach or numerical Quantum Monte Carlo simulations.

Furthermore, we have found that the temperature, the DDI and the disorder potential significantly affect the static structure factor. 
This latter provides important information on the impact of disorder and the DDI interaction on the BEC coherence.
For small values of the roton gap, $S(k)$ becomes striking leading to large density fluctuations. 
We have pointed out also that the peculiar interplay of rotonization induced by DDI and disorder may lead to strongly depress the superfluid density in the roton's region 
due to the localization of the particles in the respective minima of the external random potential.
Such an effect may open the way to new investigations of localization phenomena in quantum gases, and the superfluid behavior with varying the strength 
and the width of the disordered potential. We have also discussed the validity criterion of the Bogoliubov approximation.


Experimental studies based on our theoretical findings would be of great interest and would strengthen our analysis.
According to our estimations, the 2D dirty dipolar BECs are achievable in experiments with
ultracold atoms with highly magnetic dipolar interaction or polar molecules.

Finally, important extensions of this work concern the effect of  weak disorder in 2D dipolar gas with tilting angle where the interaction in the plane becomes anisotropic.
We also plan to investigate the localization properties of Bogoliubov quasiparticles in dipolar Bose gases.

\section{Acknowledgments}
We thank Axel Pelster for useful comments and correspondence.  
We also acknowledge stimulating discussions with Cord M\"uller and Laurent Sanchez-Palencia.


\begin{thebibliography}{28}
\bibitem{Baranov} See for review: M. A. Baranov, Physics Reports {\bf 464}, 71 (2008).
\bibitem{Pfau} See for review: T. Lahaye et al., Rep. Prog. Phys. {\bf 72}, 126401 (2009).
\bibitem{Carr} See for review: L.D. Carr, D. DeMille, R.V. Krems, and J. Ye, New Journal of Physics {\bf 11}, 055049 (2009).
\bibitem{Pupillo2012} See for review: M.A. Baranov, M. Delmonte, G. Pupillo, and P. Zoller, Chemical Reviews, {\bf 112}, 5012 (2012).
\bibitem{boudjGora}  Abdel\^{a}ali Boudjemaa and G.V. Shlyapnikov, Phys. Rev. A {\bf 87}, 025601 (2013).
\bibitem{gora} L. Santos, G.V. Shlyapnikov, and M. Lewenstein, Phys.Rev. Lett. {\bf 90}, 250403 (2003).
\bibitem{prok} See for review: M. Boninsgni and N.V. Prokof'ev, Rev. Mod. Phys. {\bf 84}, 759 (2012).
\bibitem{petGora} Zhen-Kai Lu, Yun Li, D. S. Petrov, G. V. Shlyapnikov, Phys. Rev. Lett. {\bf 115}, 075303 (2015).
\bibitem{Uwe} U.R. Fischer, Phys. Rev. A {\bf 73}, 031602(R) (2006).
\bibitem{Ron} S. Ronen, D. C. E. Bortolotti, and J. L. Bohn, Phys. Rev. Lett. {\bf 98}, 030406 (2007); S. Ronen and J. L. Bohn, Phys. Rev. A {\bf 76}, 043607 (2007).
\bibitem{bohn1} R.M. Wilson, S. Ronen, J.L. Bohn, and H. Pu, Phys. Rev. Lett. {\bf 100}, 245302 (2008).
\bibitem{Lu}  H.-Y. Lu, H. Lu, J.-N. Zhang, R.-Z. Qiu, H. Pu, and S. Yi, Phys. Rev. A {\bf 82}, 023622 (2010).
\bibitem{Wils}  R.M.Wilson, C. Ticknor, J. L. Bohn, and E. Timmermans, Phys. Rev. A {\bf 86}, 033606 (2012).
\bibitem{Mart}  A. D. Martin and P. B. Blakie, Phys. Rev. A {\bf 86}, 053623 (2012).
\bibitem{Tic}  C. Ticknor, R. M. Wilson, and J. L. Bohn, Phys. Rev. Lett. {\bf 106}, 065301 (2011).
\bibitem{Nath}  R. Nath and L. Santos, Phys. Rev. A {\bf 81}, 033626 (2010).
\bibitem{Biss}  R. N. Bisset and P. B. Blakie, Phys. Rev. Lett. {\bf 110}, 265302 (2013).
\bibitem{Land} L. D. Landau, J. Phys. (USSR), {\bf5}, 71 (1941).
\bibitem{Land1} L.D. Landau, J. Phys. (USSR), {\bf11}, 91 (1947).
\bibitem{Feym} R. P. Feynman, Phys. Rev. {\bf 94}, 262 (1954).
\bibitem{Feym1} R. P. Feynman and Michael Cohen, Phys. Rev. {\bf102}, 1189 (1956).
\bibitem{Fedo} A. K. Fedorov, I. L. Kurbakov, and Yu. E. Lozovik, Phys. Rev. B {\bf 90}, 165430 (2014).
\bibitem{mora} C. Mora et al., Phys. Rev. B {\bf 76}, 064511 (2007).
\bibitem{bush} H. P. B\"{u}chler et al., Phys. Rev. Lett. {\bf 98}, 060404 (2007).
\bibitem{Astr} G. E. Astrakharchik et al., Phys. Rev. Lett. {\bf 98}, 060405(2007).
\bibitem{prok3} A. Filinov, N.V. Prokofev, and M. Bonitz, Phys. Rev. Lett. {\bf 105}, 070401 (2010).
\bibitem {Blak2} P. B. Blakie, D. Baillie, and R. N. Bisset, Phys. Rev. A {\bf 88}, 013638 (2013).
\bibitem{San} L. Sanchez-Palencia and M. Lewenstein, Nature Phys. {\bf 6}, 87 (2010).
\bibitem{Deis} B. Deissler, M. Zaccanti, G. Roati, C. D’Errico, M. Fattori, M. Modugno, G. Modugno, and M. Inguscio, Nature Phys. {\bf 6}, 354 (2010).
\bibitem{Ale} I. L. Aleiner, B. L. Altshuler and G. V. Shlyapnikov, Nature Phys. {\bf 6}, 900 (2010).
\bibitem{Giam} T. Giamarchi and H. J. Schulz, Phys. Rev. B {\bf 37}, 325 (1988).
\bibitem{Fisher} M. P. A. Fisher, P. B. Weichman, G. Grinstein, and D. S. Fisher, Phys. Rev. B {\bf 40}, 546 (1989).
\bibitem{And} P.W. Anderson, Phys. Rev. {\bf 109}, 1492 (1958).
\bibitem{Sanch} L. Sanchez-Palencia, D. Cl\'ement, P. Lugan, P. Bouyer, G. V. Shlyapnikov, and A. Aspect, Phys. Rev. Lett. {\bf 98}, 210401, (2007).
\bibitem{Asp}  J. Billy, V. Josse, Z. Zuo, A. Bernard, B. Hambrecht, P. Lugan, D. Cl\'ement, L. Sanchez-Palencia, P. Bouyer, A. Aspect, Nature {\bf 453}, 891 (2008).
\bibitem{Asp1}  G. Roati, C. D’Errico, L. Fallani, M. Fattori, C. Fort, M. Zaccanti, G. Modugno, M. Modugno, M. Inguscio, Nature {\bf 453}, 895 (2008).

\bibitem{Huang} K. Huang and H. F. Meng, Phys. Rev. Lett. {\bf 69}, 644 (1992).
\bibitem{AxelM} Mahmoud Ghabour and Axel Pelster, Phys. Rev. A {\bf 90}, 063636 (2014).
\bibitem{Boudj} Abdel\^{a}ali Boudjem\^{a}a, Phys. Rev. A {\bf 91}, 053619 (2015).
\bibitem{Boudj1} Abdel\^{a}ali Boudjem\^{a}a, J. Low Temp. Phys. {\bf 180},  377 (2015).
\bibitem{Axel1} Christian Krumnow and Axel Pelster, Phys. Rev. A {\bf 84}, 021608(R) (2011).
\bibitem{Axel2} Branko Nikolic, Antun Balaz, Axel Pelster, Phys. Rev. A {\bf 88}, 013624 (2013).
\bibitem{Boudj2} Abdel\^{a}ali Boudjem\^{a}a, Phys.Lett.A, {\bf 379} 2484 (2015).


\bibitem{Mich}  Michikazu Kobayashi and Makoto Tsubota, Phys. Rev. B {\bf 66}, 174516 (2002).
\bibitem{Sanch1} L. Sanchez-Palencia, Phys. Rev. A {\bf 74}, 053625 (2006).
\bibitem{Allard} B. Allard, T. Plisson, M. Holzmann, G. Salomon, A. Aspect, P. Bouyer, and T. Bourdel, Phys. Rev. A {\bf 85}, 033602 (2012).
\bibitem{Sal} Joseph Saliba, Pierre Lugan, Vincenzo Savona, Phys. Rev. A {\bf 90}, 031603 (R) (2014).
\bibitem{Lev} L. Pitaevskii and S. Stringari, Phys. Rev. Lett. {\bf8}1, 4541 (1998).
\bibitem{boudj2015} Abdel\^{a}ali Boudjem\^{a}a, J. Phys. B: At. Mol. Opt. Phys. {\bf 48} 035302 (2015).
\bibitem{Gior}  S. Giorgini, L. Pitaevskii, and S. Stringari, Phys. Rev. B {\bf 49}, 12 938 (1994).


\bibitem{Fal} G. M. Falco, A. Pelster, and R. Graham, Phys. Rev. A {\bf 75}, 063619 (2007).
\bibitem{Gaul} C. Gaul, N. Renner, and C. A. M\"uller, Phys. Rev. A {\bf 80}, 053620 (2009).
\bibitem{Gaul1} C. Gaul and C. A. M\"uller, Phys. Rev. A {\bf 83}, 063629 (2011).
\bibitem{Yuk} V. I. Yukalov and R. Graham, Phys.Rev.A {\bf 75}, 023619 (2007).

\bibitem{merm} N. D. Mermin, and H. Wagner, Phys. Rev. Lett. {\bf 22}, 1133 (1966).
\bibitem{hoh} P. C. Hohenberg, Phys. Rev. {\bf 158}, 383 (1967).
\bibitem{KT} J.M. Kosterlitz and D.J. Thouless, J.Phys. C {\bf 6}, 1181 (1973); J.M. Kosterlitz, J. Phys. C {\bf 7}, 1046 (1974).
\bibitem{pop} V.N. Popov, {\it Functional Integrals in Quantum Field Theory and Statistical Physics} (D. Reidel Pub., Dordrecht, 1983).
\bibitem{petr1} D. S. Petrov, M. Holtzmann, and G. V. Shlyapnikov,Phys. Rev. Lett. {\bf 84}, 2551 (2000).
\bibitem{boudj2012} Abdel\^{a}ali Boudjem\^{a}a, Phys. Rev. A {\bf 86}, 043608 (2012).
\bibitem{GPS} D.S. Petrov, D.M. Gangardt, and G.V. Shlyapnikov, J. Phys. IV (France) {\bf 116}, 5 (2004).
\bibitem{Noz} P. Nozi\`{e}res, J. Low Temp. Phys. {\bf 142}, 91 (2006); {\it ibid} {\bf 156}, 9 (2009).  
\bibitem{bali} See for review: S. Balibar, A.D. Fefferman, A. Haziot, and X. Rojas, J. Low Temp. Phys. {\bf 168}, 221 (2012).

\bibitem{Lopa} A.V. Lopatin and V. M. Vinokur, Phys. Rev. Lett. {\bf 88}, 235503 (2002).

\bibitem{Lug} P. Lugan, D. Cl\'ement, P. Bouyer, A. Aspect, and L. Sanchez-Palencia, Phys. Rev. Lett. {\bf 99}, 180402 (2007).
\bibitem{Mul}  C.A. M\"uller, Phys. Rev. A {\bf 91}, 023602 (2015).
\bibitem{Gaul3} C. Gaul, P. Lugan, C.A. M\"uller,  Ann. Phys. (Berlin) {\bf 527}, 531 (2015).

\end{thebibliography}
\end{document}